\documentclass[journal]{IEEEtran}
\usepackage{amsmath}
\usepackage{amsfonts}
\usepackage{graphicx}
\usepackage{subfigure}
\usepackage{epstopdf}
\usepackage{algorithm}
\usepackage{algorithmic}
\usepackage{lineno,hyperref}
\usepackage{mathrsfs}
\usepackage{nomencl}
\usepackage{etoolbox}
\usepackage{multirow}
\usepackage{bigstrut}
\usepackage{float}
\usepackage{color}
\usepackage{bm}
\renewcommand\nomgroup[1]{
	\item[\bfseries
	\ifstrequal{#1}{A}{Sets}{
		\ifstrequal{#1}{B}{Parameters}{%
			\ifstrequal{#1}{C}{Variables}{%
				\ifstrequal{#1}{D}{Indices}{}}}}%
	]}
\makenomenclature 
\begin{document}
	\title{Smart Online Charging Algorithm for Electric Vehicles via Customized Actor-Critic Learning
		\thanks{\textcolor[rgb]{0,0,0}{Y. Cao, D. Li, and G. Zhang are with College of Information Science and Technology, Donghua University; Engineering Research Center of Digitized Textile and Apparel Technology, Ministry of Education, Shanghai 201620, China (Email: yongshengcao@mail.dhu.edu.cn, deminli@dhu.edu.cn, glzhang@dhu.edu.cn). H. Wang is with the Department of Data Science and Artificial Intelligence, Faculty of Information Technology, Monash University, Melbourne VIC 3800, Australia, and also with the Stanford Sustainable Systems Lab, Stanford University, Stanford, CA 94305 USA  (Email: hao.wang2@monash.edu). }This work is supported by the National Natural Science Foundation of China (Grant No. 61772130, 61301118); the International Science and Technology Cooperation Program of the Shanghai Science and Technology Commission (Grant No. 15220710600); the Innovation Program of the Shanghai Municipal Education Commission (Grant No. 14YZ130); China Scholarship Council (File No. 201906630026); the Fundamental Research Funds for the Central Universities (Grant: CUSF-DH-D-2018093); the FIT Academic Staff Funding of Monash University. (Corresponding author: Guanglin Zhang, Hao Wang.)}
	}
	\author{Yongsheng Cao, Hao Wang, \textit{Member, IEEE}, Demin Li, Guanglin Zhang, \textit{Member, IEEE}}
	\maketitle
	\begin{abstract}
		With the advances in the Internet of Things technology, electric vehicles (EVs) have become easier to schedule in daily life, which is reshaping the electric load curve. It is important to design efficient charging algorithms to mitigate the negative impact of EV charging on the power grid. This paper investigates an EV charging scheduling problem to reduce the charging cost while shaving the peak charging load, under unknown future information about EVs, such as arrival time, departure time, and charging demand. First, we formulate an EV charging problem to minimize the electricity bill of the EV fleet and study the EV charging problem in an online setting without knowing future information. We develop an actor-critic learning-based smart charging algorithm (SCA) to schedule the EV charging against the uncertainties in EV charging behaviors. \textcolor[rgb]{0,0,0}{The SCA learns an optimal EV charging strategy with continuous charging actions instead of discrete approximation of charging.} We further develop a more computationally efficient customized actor-critic learning charging algorithm (CALC) by reducing the state dimension and thus improving the computational efficiency. Finally, simulation results show that our proposed SCA can reduce EVs' expected cost by $24.03\%$, $21.49\%$, $13.80\%$, compared with the Eagerly Charging Algorithm, Online Charging Algorithm, RL-based Adaptive Energy Management Algorithm, respectively. CALC is more computationally efficient, and its performance is close to that of SCA with only a gap of $5.56\%$ in the cost.
	\end{abstract}
	\nomenclature[A]{$\mathcal{T}$}{Time horizon.}
	\nomenclature[A]{$\mathcal{N}$}{Set of EVs.}
	\nomenclature[B]{$k_0,k_1$}{Coefficients of the electricity price model.}
	\nomenclature[B]{$b_{\max}$}{Maximum charging amount in a time slot.}
	\nomenclature[B]{$\gamma$}{Discount factor.}
	\nomenclature[B]{$\beta_a$}{Learning rate for the actor process.}
	\nomenclature[B]{$\beta_c$}{Learning rate for the critic process.}
	\nomenclature[C]{$l_{ev}(t)$}{Charging load of the EV fleet.}
	\nomenclature[C]{$b_i(t)$}{Charging amount in time slot $t$ of EV $i$.}
	\nomenclature[C]{$t_i^{arr}$}{Arrival time of EV $i$.}
	\nomenclature[C]{$t_i^{dep}$}{Departure time of EV $i$.}
	\nomenclature[C]{$\hat{D}_i(t)$}{Electricity demand of EV $i$ in time slot $t$.}
	\nomenclature[C]{$\mathcal{H}(t)$}{Set of EVs that are parked in the charging station in time slot $t$.}
	\nomenclature[C]{$\mathcal{W}(t)$}{Set of the rolling window from the current time slot $t$ to $t'$.}
	\nomenclature[C]{$l_b(t)$}{Inelastic base load of the other electricity demand from the community in time slot $t$.}
	\nomenclature[C]{$SOC_{i,t}$}{State of charge (SOC) of EV $i$ in time slot $t$.}
	\nomenclature[C]{$p(t)$}{Electricity price in time slot $t$.}
	\nomenclature[C]{$r(t)$}{Reward function in time slot $t$.}
	\nomenclature[C]{$\phi(t)$}{State in time slot $t$.}
	\nomenclature[C]{$\pi_\theta(\cdot)$}{Gradient policy.}
	\nomenclature[C]{$\theta$}{Thread parameter.}
	\nomenclature[D]{$t$}{Index of time slot.}
	\nomenclature[D]{$i$}{Index of EV.}
	\nomenclature[D]{$k$}{Index of global shared counter.}
	\printnomenclature 
	\section{Introduction}
	With the increasing concerns of environmental issues, electric vehicles (EVs) emerge as a promising solution as they do not directly consume fossil fuels and are more environmentally friendly. Meanwhile, the intermittent charging demands caused by electric vehicles (EVs) impact the operation of the public power grid \cite{IntegrationofElectricVehicles}. Therefore, it is crucial to design charging control strategies to alleviate the peak load caused by EVs and cut down their electricity bills. It will be ideal if the future charging demand is known in advance, such that the EV charging can be scheduled to flatten the total load \cite{OptimalSchedulingforChargingandDischarging}. However, an EV charging station faces great uncertainties in EVs' behaviors, including their travel patterns and charging demands. Online charge strategies become a promising paradigm for determining the optimal charging of EVs against uncertainties. \textcolor[rgb]{0,0,0}{The online EV charging problem is more practical, as it does not assume any future information but only relies on the current and past EV profiles, including the arrival time, the departure time, and the charging demand of EVs.} The advances in the Internet of Thing (IoT) technology and the intelligent transportation system have paved the way for EVs \cite{Hu2019}. The information shared between EVs can improve real-time transportation and make smart decisions for individual EVs \cite{Lu2014}. It is easier to predict EVs' behaviors that make it possible to schedule a proper quantity of EVs to make charging decisions. In this paper, we aim to design online algorithms for EV charging without knowing future information.

	In recent years, great efforts have been made to develop online EV charging algorithms under stochastic EV demands. Online charging problems do not assume future information about the profiles of EVs, which captures the realistic scenarios about uncertainties in EV charging behaviors. Yu \textit{et al}. \cite{DistributedOnlineEnergyManagement} proposed a distributed online algorithm by the Lyapunov optimization method and an improved alternating direction method to investigate an energy scheduling problem for distributed data centers and EVs. Qi \textit{et al}. \cite{DevelopmentandEvaluationofanEvolutionary} proposed an online energy management framework of EVs by an evolutionary algorithm. 
	\textcolor[rgb]{0,0,0}{Kang \textit{et al}. \cite{Kang2016} presented a novel centralized EV charging strategy based on spot price with the consideration of charging priority and charging location.}
	Li \textit{et al}. \cite{Li2019} designed a joint online learning and pricing algorithm to minimize the operational cost of utility considering time-varying demand responses and consumers' responses. 
	\textcolor[rgb]{0,0,0}{A novel multi-objective evolutionary algorithm was proposed in \cite{Kang2017} to minimize the peak-to-valley difference of the load and the operating cost.}
	Quddus \textit{et al}. \cite{QUDDUS2018841} proposed a two-stage stochastic programming model to optimize the power flow of commercial buildings and EV charging stations with some practical constraints.
	However, these algorithms \cite{DistributedOnlineEnergyManagement}-\cite{QUDDUS2018841} relied on specific models or they only worked in special scenarios. Instead, we aim to develop a generic method for the EV charging problem that is less model-dependent and can work for various practical scenarios. Therefore, we refer to the model-free reinforcement learning (RL) to derive the optimal EV charging strategy in our work.

	Model-free RL frameworks and policies have been utilized to handle the energy scheduling problems in the literature. For example, an improved Q-learning method has been proposed to minimize the electricity bill and reduce users' discomfort for a household \cite{StochasticControlforSmartGrid}\cite{OptimalDemandResponse}. \textcolor[rgb]{0,0,0}{A bidirectional long short-term memory network-based parallel reinforcement learning was presented in \cite{Liu2020} to construct an energy management strategy for a hybrid tracked vehicle.} A control algorithm based on Q-learning has been designed in \cite{Duan2018} to obtain the optimal control under physical and cyber uncertainties. A batch RL algorithm has been investigated in \cite{ResidentialDemandResponseBatch} to schedule controllable load such as washing machine. However, these studies \cite{StochasticControlforSmartGrid}-\cite{ResidentialDemandResponseBatch} all used the RL methods, in which the charging actions have to be discrete values that restrict the model of actions. \textcolor[rgb]{0,0,0}{Using discrete actions for EV charging is an approximation to the real-world problem, as the EV charging amount is a continuous value. When the discrete-charging action is adopted, it is often difficult to achieve a good trade-off between computational complexity and performance. Using too few discrete levels may result in poor learning performance, but using a large discrete action space will make the training difficult and lead to high computation overhead \cite{Hwang2013}.} 
	
	In our work, we aim to develop optimal EV charging strategies with continuous charging amount other than discretize charging actions. The probability distribution of actions under different states is a stochastic policy, and deriving an optimal policy is at the center of RL methods. A new method was discussed in \cite{Bertsekas2019} to approximate the iteration of policy and improve the policy. Standard tabular Q-learning or deep Q neural network method can not derive the optimal policy over continuous action space \cite{Jiang2019}\cite{Bui2020}. Therefore, we adopt the actor-critic method \cite{sutton} to solve our problem with continuous states and actions. The actor-critic method can use the deep neural network (DNN) to estimate the value functions, which can get a better approximation. There are some studies on the scheduling problem using the actor-critic method. \textcolor[rgb]{0,0,0}{Some recent studies in \cite{Kamalapurkar2014}-\cite{Wang2021} developed actor-critic learning approaches for various applications. For example, a concurrent actor-critic learning framework was proposed in \cite{Kamalapurkar2014} to achieve a close-to-optimal feedback-Nash equilibrium solution to a multi-player non-zero-sum differential game in an infinite horizon. A distributed framework based on policy search was proposed in \cite{Cao2019} to accelerate the learning processes of robot moving by reducing variance. An actor-critic approach was proposed in \cite{Wang2021} to approximate the performance function based on adaptive dynamic programming strategy.} Lu \textit{et al}. \cite{LU2019937} proposed a real-time incentive-based algorithm to help the service provider balance energy fluctuations and improve the reliability of smart grid systems using deep RL. 
	Wei \textit{et al}. \cite{UserSchedulingandResourceAllocation} proposed a policy-gradient method to study the problem of user scheduling and resource allocation in heterogeneous networks with continuous states and actions. However, the actor-critic algorithm may not have a good convergence if its policy is on-policy \cite{David2014}. \textcolor[rgb]{0,0,0}{Hence, we design an online EV charging strategy using asynchronous actor-critic learning, and we further develop a customized actor-critic algorithm, significantly improving the convergence and achieving a close-to-optimal solution of the charging schedule of EV fleet.}

	This paper aims to design an online EV charging scheduling algorithm leveraging the EV charging data. \textcolor[rgb]{0,0,0}{We first formulate an offline optimization problem, in which the future profiles of EVs are known. The offline problem captures all the modeling components, including energy cost, arrival time, departure time, and the charging demand of EVs, thus serving as a basic formulation for EV charging. As the offline problem is unrealistic, we further formulate an online optimization problem for EV charging without assuming future profiles of EVs and develop online charging strategies.} We model the EV charging decision as a Markov decision process (MDP), where the charging station determines the charging schedule according to the past and current information, including the arrival time, departure time, and charging demand of EVs. In the MDP setup, the current charging decision will affect the next state, the charging decision, and accumulative rewards in the future. We aim to develop an optimal EV charging strategy to minimize the expected total energy cost under uncertainties of EV charging behaviors. We develop an actor-critic learning-based smart charging algorithm (SCA), which determines the optimal continuous-charging amount for each EV using asynchronous actor-critic reinforcement learning. To further improve the computational efficiency of SCA, we develop a customized \textcolor[rgb]{0,0,0}{actor-critic learning} charging algorithm (CALC) that reduces the dimension of the state during the learning phase. Finally, SCA and CALC are compared with three state-of-the-art algorithms. We summarize the contributions of our paper as follows.
	\begin{itemize}
	\item 	\textcolor[rgb]{0,0,0}{We model the online EV charging problem as a Markov decision process to capture the decision marking under uncertainty of EV charging profiles.} We develop a smart charging algorithm (SCA) to solve the online EV charging problem, which leverages the advantage of the asynchronous actor-critic method with good convergence to derive the optimal charging policy.
		\item 	\textcolor[rgb]{0,0,0}{We further develop a more computationally efficient customized actor-critic learning charging algorithm (namely CALC), which consists of two stages. In the first stage, CALC learns the charging policy for the whole group of EVs using actor-critic learning and obtains the optimal aggregate charging amount; in the second stage, CALC allocates the aggregate charging amount to serve each EV based on the projection theory. Such a customized algorithm achieves sub-optimal performance and significantly reduces the computational overhead, achieving a good trade-off between performance and computation.}
		\item 	\textcolor[rgb]{0,0,0}{The developed SCA and CALC algorithms learn EV charging strategies with continuous charging actions instead of discrete approximation of charging. We compare our developed charging algorithms with Q-network-based RL algorithm, namely adaptive energy management (AEM), which makes discrete actions. Our results show that our developed charging algorithms outperform AEM with different numbers of discretized actions in achieving a better trade-off between computation and performance.}
	\end{itemize}
	The remainder of this paper consists of five sections. Section \ref{sectionII} formulates the offline EV scheduling problem. Section \ref{sectionIII} extends the offline EV charging problem to an online charging scheduling problem. Then SCA and CALC algorithms are proposed and analyzed in Section \ref{sectionIV}. Simulation results are presented in Section \ref{sectionV}. Finally, Section \ref{sectionVI} concludes the paper. \textcolor[rgb]{0,0,0}{The summary of notations is shown in Nomenclature.}

	\section{Offline EV charging problem} \label{sectionII}
	In this section, we investigate the optimal offline EV charging problem, where the future information of EVs is known in advance. \textcolor[rgb]{0,0,0}{The offline EV charging optimization problem will help formulate the online charging problem in Section \ref{sectionIII}.}
	\subsection{System Architecture}
	We consider a community including an EV charging station and the inelastic base load. 
	\textcolor[rgb]{0,0,0}{The charging station serves EVs in a region, and we restrict our discussion for the charging management at one charging station.}
	We aim to minimize the total charging cost of EVs from the charging station. We study the charging scheduling of EV battery in a time horizon $\mathcal{T}$ and the time index $t\in \mathcal{T}=\{1,2,...,T\}$. We assume that $N$ EVs arrive in the order from $1$ to $N$ and the index of EV is $i \in \mathcal{N}=\{1,2,...,N\}$. We set the arrival time and departure time of EV $i$ as $t_{i}^{arr}$ and $t_{i}^{dep}$, respectively. Let $D_{i}$ denote the charging demand of EV $i$. \textcolor[rgb]{0,0,0}{In the traditionally offline models, the charging station knows the EV profiles $D_{i}, t_{i}^{arr}$, and $t_{i}^{dep}$ in advance, which is unrealistic. This paper will consider an online model, in which the charging station does not know any future information about EV profiles, capturing the key uncertainty in a real-world scenario.}

	According to the physical constraints of EV battery, EV $i$ should be charged at a \textcolor[rgb]{0,0,0}{charging amount $b_{i}(t)$ in time slot $t$} and the \textcolor[rgb]{0,0,0}{charging amount in each time slot} has a bound, that is,
	\begin{align}
	b_{i}(t) \in [0, b_{i,\max}],
	\end{align}
	where $b_{i,\max}$ is the maximum \textcolor[rgb]{0,0,0}{charging amount in a time slot} of EV $i$. 
	\textcolor[rgb]{0,0,0}{We set the state of charge (SOC) of EV $i$ in time slot $t$ as $SOC_{i,t}$, which is defined as 
		$SOC_{i,t}=\frac{B_{i}(t)}{B_{i,\max}},$
		where $B_{i,\max}$ is the battery capacity of EV $i$ and the battery level follows $B_{i}(t+1)=B_{i}(t)+b_i(t)$.} We denote $\mathcal{H}(t)$ as the set of EVs that are parked in the charging station in time slot $t$.  
	The charging station can control the \textcolor[rgb]{0,0,0}{charging amount} $b_{i}(t)$. We denote the charging load of \textcolor[rgb]{0,0,0}{EV fleet} $l_{ev}(t)$ in time slot $t$, that is,
	\begin{align} 
	l_{ev}(t)=\sum_{i\in\mathcal{H}(t)}b_{i}(t).
	\end{align}
	Except for the EV charging load, we also consider the inelastic base load of the other electricity demand from the community $l_b(t)$, such as lighting or watching TV. We assume that the base load $l_b(t)$ can be predicted accurately at the beginning of each time slot $t$ and thus is known to the operator. The total load in time slot $t$ is $L(t)$ and we have
	\begin{align}
	L(t)=l_{ev}(t)+l_b(t).
	\end{align}
	In practice, \textcolor[rgb]{0,0,0}{the total load includes the total EV charging load and base load and is upper-bounded by $L_{\max}$.}
	According to \cite{DecentralizedChargingControlofLargePopulationsofPlug-inElectricVehicles}, the unit electricity price $p(t)$ \textcolor[rgb]{0,0,0}{is} modeled as a linear function of the total load,
	\begin{align}\label{eprice}
	p(t)=k_{0}+2k_{1}(l_{ev}(t)+l_b(t)),
	\end{align}
	where $k_{0}$ and $k_{1}$ are non-negative coefficients. \textcolor[rgb]{0,0,0}{Essentially, reducing the electricity cost is to shift the load evenly and avoid significant peak load.} We denote $\boldsymbol{b}(t)$ as a vector form of $(b_1(t), b_2(t), ..., b_i(t))$
	and calculate the electricity bill $c(\boldsymbol{b}(t))$ of the charging station as follows,
	\begin{align}\label{eqn_price}
	c(\boldsymbol{b}(t))&=\int_{l_b(t)}^{L(t)}(k_{0}+2k_{1}z)d z\nonumber\\ 
	&=k_{0}\sum_{i\in\mathcal{H}(t)}b_{i}(t)+k_{1}\Big(\sum_{i\in\mathcal{H}(t)}b_{i}(t)\Big)^{2}\nonumber\\&+2k_{1}l_b(t)\sum_{i\in\mathcal{H}(t)}b_{i}(t),
	\end{align}
	where the electricity bill $c(\boldsymbol{b}(t))$ is the integral of the unit electricity price $p(t)$ in the load interval from $l_b(t)$ to $L(t)$.
	%
	
	\subsection{Problem Formulation}
	We first formulate the offline optimization problem assuming that we know the EV arrival time, departure time, and charging demand. We aim to find the optimal charging solution $b_i(t)$ of each EV $i$ to minimize the total EV charging cost in an operational horizon $\mathcal{T}$, where the profiles of EVs are known ahead.
	Then, we can formulate the offline charging optimization problem as follows,
	\begin{subequations} \label{cost_problem} 
		\begin{align}
		&\textcolor[rgb]{0,0,0}{  \min_{b_i(t)}   \sum_{t=1}^{T}\Big(k_{0}\sum_{i\in\mathcal{N}}b_{i}(t)+k_{1}\Big(\sum_{i\in\mathcal{N}}b_{i}(t)\Big)^{2}  }\nonumber\\
		&\hspace{6mm}+2k_{1}l_b(t)\sum_{i\in\mathcal{N}}b_{i}(t)\Big) \\
		&\mathrm{s.t.} \sum_{t_{i}^{arr}}^{t_{i}^{dep}} b_{i}(t)=D_{i}, i\in \mathcal{N} \\
		&\hspace{6mm} 0 \le b_{i}(t) \le b_{i,\max}, i\in \mathcal{N}, t\in \mathcal{T}\\
		&\hspace{6mm} 0 \le L(t) \le L_{\max}, t\in \mathcal{T}
		\end{align}
	\end{subequations}
	where $D_{i}$ is the charging demand of EV $i$, $t_{i}^{arr}$ is the arrival time of EV $i$, $t_{i}^{dep}$ is the departure time of EV $i$. We can find that problem (\ref{cost_problem}) is a convex optimization problem. If the profiles $t_{i}^{arr}, t_{i}^{dep}$ and $D_{i}$ are known in advance, the optimal charging solution $b_{i}(t)$ can be attained by solving the optimization problem (\ref{cost_problem}). Nevertheless, the current EV charging information, including the arriving time $t_{i}^{arr}$, the departure time $t_{i}^{dep}$ and $D_{i}(t)$ can only be known when it arrives at the charging station. In the next section, we will study an online EV charging problem derived from this offline problem, and the future information of EVs is unknown in the online charging problem.

	\section{Online EV Charging Problem} \label{sectionIII}
	In this section, we reformulate an online charging problem based on \eqref{cost_problem}. 
	We denote $\mathcal{Q}(i,t_{s})$ as the set of time indices that EV $i$ will charge in the charging station, \textcolor[rgb]{0,0,0}{and} $\mathcal{Q}(i,t_{s})=[t_s,t_{i}^{dep}]$. The \textcolor[rgb]{0,0,0}{charging amount in time slot $t$} of EV $i$ is defined as $b_{i}(t)$, where $t \in \mathcal{Q}(i,t_{s})$.
	When an EV comes to the charging station in time slot $t_{s}$, we optimize the charging scheduling without knowing the demand of EVs in the future. 
	
	In the online charging problem, the profiles of EVs including the charging demand, the arrival and departure time, are not known ahead. \textcolor[rgb]{0,0,0}{The charging station needs to schedule the EV charging for each current time slot while facing uncertainties of EV profiles in the future.} According to \cite{powell2020state}, we define the exogenous information variable $\xi(t)$ that comes up in time slot $t$, which is not known ahead when the charging decision $\boldsymbol{b}(t)$ is made. We have the exogenous information variables $\xi(t)$,
	\begin{align}\label{xiexo}
	\xi(t)=(\hat{t}_{i}^{arr},\hat{t}_{i}^{dep},\hat{D}_{i}(t)),
	\end{align}	
	where $\hat{t}_{i}^{arr}, \hat{t}_{i}^{dep},\hat{D}_{i}(t)$ are the arrival time, departure time and the electricity demand of EV $i$ in time slot $t$. \textcolor[rgb]{0,0,0}{These exogenous variables could be field observations \cite{powell2020state}, bringing uncertainties and challenges to the problem solving.} We aim to find the optimal charging solution $\boldsymbol{b}=(\boldsymbol{b}(1),...,\boldsymbol{b}(T))$ to minimize the total EV charging cost $C_T$ in an operational horizon $\mathcal{T}$,
	\begin{align}
	C_T=\sum_{t=1}^{T}c(\boldsymbol{b}(t)),
	\end{align}
	where $T$ is ending time of the period $\mathcal{T}$. Therefore, we have an online charging optimization problem as follows, 
	\begin{subequations} \label{cost_problem_oa_discrete}
		\begin{align}
		&\min_{\boldsymbol{b}(t)}   \mathbb{E}[C_T] \\
		&\mathrm{s.t.} \sum_{t\in \hat{\mathcal{Q}}(i,t_{s})} b_{i}(t)=\hat{D}_i(t_s),i\in \mathcal{H}(t_{s})\\
		&\hspace{6mm} 0 \le b_{i}(t) \le b_{i,\max}, i\in \mathcal{H}(t_{s}), t\in \hat{\mathcal{Q}}(i,t_{s}),\\
		&\hspace{6mm} 0 \le L(t) \le L_{\max}, t\in \hat{\mathcal{Q}}(i,t_{s}),
		\end{align}
	\end{subequations}
	where $\hat{\mathcal{Q}}(i,t_s)=[t_s, \hat{t}_{i}^{dep}]$, and \textcolor[rgb]{0,0,0}{$\mathcal{H}(t_s)$ is the set of EVs that \textcolor[rgb]{0,0,0}{park in the charging station} in time slot $t_s$ and will remain in the charging station at time $t$, $t\in \mathcal{Q}(i,t_{s})$.
		We define $\mathcal{W}(t)$ as the set of the rolling window from the current time slot $t$ to $t'$, where $t'$ is the maximum departure time of EVs in $\mathcal{H}(t)$, when EVs are parked in the charging station, that is,
		\begin{align} \mathcal{W}(t)=\{t'|t' \ge t \:\& \: t' \le \max\{t_{i}|i \in \mathcal{H}(t)\}\}.\end{align}}
	\begin{figure}[htb] 
		\centering
		\includegraphics[width=0.4\textwidth]{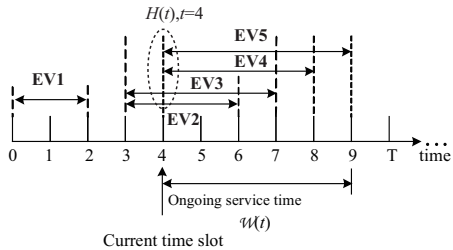}
		\vspace{-2mm}
		\caption{Illustration of $\mathcal{H}(t)$ and $\mathcal{W}(t)$.}
		\label{model_tpc2}
	\end{figure}
	\textcolor[rgb]{0,0,0}{Fig. \ref{model_tpc2} shows an example to explain the concept of $\mathcal{H}(t)$ and $\mathcal{W}(t)$. There are five EVs at the charging station in this example. In time slot $4$, there are four EVs in this charging station, e.g., $\mathcal{H}(4)=\{2,3,4,5\}$. In this time slot, the maximum service time of EVs parking in this charging station is from $t=4$ to $t'=9$, e.g., $\mathcal{W}(4)=\{4,5,6,7,8,9\}$ and EV $5$ is the last one leaving the charging station in time slot $9$.
		For the rolling-based method according to \cite{zivot2007modeling}, we replace the interval $t\in \hat{\mathcal{Q}}(i,t_{s})$, in which EV $i$ stays in the charging station as $t\in \mathcal{W}(t_{s})$.}   \textcolor[rgb]{0,0,0}{Heuristic rolling-based online control is a general method widely used in various problems of smart grids, such as real-time energy scheduling \cite{Gong2013} and ancillary services \cite{Karagiannopoulos2020}.} The charging station will implement the optimal solution $opt[b_{i}(t)]$ to Problem  (\ref{cost_problem_oa_discrete}) until a new EV comes in. When a new EV arrives, or EV completes charging, or the load from the community changes, $\mathcal{H}(t_{s}), \hat{\mathcal{Q}}(i,t_{s}), \hat{D}_i(t_s)$ should be updated and Problem (\ref{cost_problem_oa_discrete}) should be solved again.
	
	\textcolor[rgb]{0,0,0}{In practice, the offline problem is not practical because we do not have future information about EV profiles, and it is often challenging to obtain reliable prediction of human behaviors.} 
	\textcolor[rgb]{0,0,0}{In the online charging setting, it is difficult to solve this optimization problem using an intuitive mathematical programming approach, and the aforementioned rolling-based algorithm is heuristic and not optimal. We only use the rolling-based online control algorithm (OA) as a benchmark for our proposed algorithms in Section \ref{sectionIV}.}
	

	\section{RL-based EV Charging Algorithm}  \label{sectionIV}
	To tackle the challenges brought by the uncertainty of the EV behaviors, we seek an intelligent EV charging strategy using reinforcement learning.
	Considering that EV charging amount is continuous, we apply the actor-critic algorithm to solve the online EV charging problem, which combines the value-based and policy-based method.

	\subsection{RL Framework Formulation}
	We model EV charging decision as a Markov decision process. \textcolor[rgb]{0,0,0}{The charging station will make the charging schedule based on the past and current exogenous information, which includes the arrival time and departure time, and charging demand of EVs that stay in the charging station at the current time slot.} The state $\phi(t)$ consists of electricity price and the SOC of EV $i$. We measure $SOC_i(t)$ as the percentage of the battery capacity $B_{\max}$ of EV $i$. Then we have the state $\phi(t) \in \Phi$ as follows,
	\begin{align}
	\phi(t)=(SOC_1(t),...,SOC_N(t),p(t)),
	\end{align}
	where $SOC_i(t)$ is the SoC of EV $i\in\mathcal{H}(t)$ and $p(t)$ is the electricity price. The electricity price is modeled as a linear function of total load in the community so that base load can influence the price. 
	
	The action \textcolor[rgb]{0,0,0}{consists of all} the charging amount $b_i(t)$ of EV $i$ in time slot $t$, \textcolor[rgb]{0,0,0}{which are continuous variables,
		\begin{align}
		b_i(t)\in [0, b_{i,\max}], i\in \mathcal{H}(t_{s}), t\in \hat{\mathcal{Q}}(i,t_{s}),
		\end{align} 
		where $b_{i,\max}$ is the maximum charging amount of EV $i$ in each time slot, and $b_i(t)$ \textcolor[rgb]{0,0,0}{should} be in the range $[0, b_{i,\max}]$}. The charging action needs to satisfy the following constraint,
	\begin{align}
	\sum_{t=t_s}^{\hat{t}_{i}^{dep}}b_i(t)=\hat{D}_i(t)-\sum_{t=\hat{t}_i^{arr}}^{t_s-1} \tilde{b}_i(t),
	\end{align}
	where $t_s$ is the current time slot, $\hat{D}_{i}(t)$ is the electricity demand of EV $i$ with exogenous information in time slot $t$, and $\sum_{t=\hat{t}_i^{arr}}^{t_s-1}$ is the actual \textcolor[rgb]{0,0,0}{charging amount} of EV $i$ from the arrival time $\hat{t}_i^{arr}$ to the moment before the current time slot $t_s-1$.
	\textcolor[rgb]{0,0,0}{During the training, the environment $(\phi(t),b_i(t),r(t))$ is composed by the state including the base load from the community and the exogenous information (such as the arrival time $\hat{t}_{i}^{arr}$, departure time $\hat{t}_{i}^{dep}$, the EV charging demand $\hat{D}_{i}(t)$ of EV $i$ in time slot $t$), the charging action $b_i(t)$, and the reward $r(t)$. Each decision for charging action $\tilde{b}_i(t)$ will affect the state, including residual EV charging demand $\hat{D}_i(t)$ for future time slots in turn.}
	
	According to the relationship \eqref{eprice} between the electricity price and the load, we use the EV charging cost to set the reward function $r(t)$ as 
	\begin{align}
	r(t)=-\sum_{i\in\mathcal{H}(t)}(k_0+2k_1b_i(t)+2k_1l_b(t))b_i(t),
	\end{align}
	where $b_i(t)$ is the charging amount.
	To evaluate the expected accumulated rewards of current state with action $b_i(t)$ and use the policy $\pi_\theta$ to choose the charging action according to the state $\phi(t)$, we denote the state-action value function $Q^{\pi_\theta}(\phi(t),b_i(t))$ which is the value of taking the charging action $b_i(t)$ in state $\phi(t)$ under a policy $\pi_\theta$ as follows,
	\begin{align}
	Q^{\pi_\theta}(\phi(t),b_i(t))=\mathbb{E}_{\pi_\theta}\{\sum_{k=0}^{T}[\epsilon^{k}r(t+k)]|(\phi(t),b_i(t),{\pi_\theta})\},
	\end{align}
	where $\epsilon\in (0,1)$ is the discount factor and the policy $\pi_\theta$ is a function of the parameter $\theta$.
	
	Different from the state value function which is the optimal reward function according to current state and the stationary policy, such as Greedy policy, the state-action value function of actor-critic is the expected rewards according to current state. It utilizes a parameterized policy to select the charging action, which can be given by,
	\begin{equation}
	Q^{\pi}(\phi(t),b_i(t))=\mathbb{E}\{r(t)+\gamma Q^{\pi}(\phi(t+1),b_i(t+1))\},
	\end{equation}
	where $\mathbb{E}\{\cdot\}$ is the expectation function, $\gamma$ is the discount factor to evaluate foresighted decisions and $\pi$ can be approximated by $\pi_\theta(\phi(t),b_i(t))$. 
	
	\textcolor[rgb]{0,0,0}{We optimize the policy $\pi_{\theta}$ with a Gaussian distribution $\pi_{\theta}\sim\mathcal{N}(\mu_{\theta},\sigma_{\theta}^2)$, where the expectation $\mu_{\theta}$ and logarithmic standard deviation log$\sigma_{\theta}$ are approximated by the multi-layer perceptron (MLP), which can be expressed as follows,
		\begin{align}
		\mu_{\theta}&=\alpha_\mu^{\top} h +\zeta_\mu,\\
		\mathrm{log}\sigma_{\theta}&= \alpha_\sigma^{\top},
		\end{align} 	
		where $\alpha_\mu,\alpha_\sigma$ are the output layer's weights, $\zeta_\mu$ is output layer's bias and $(\cdot)^\top$ is the operation of taking the transpose. The parameter $\theta$ is the network weights of MLP and $\alpha_\mu,\alpha_\sigma, \zeta_\mu \in \theta$. The feature $h$ is extracted from the hidden layers of MLP, which can be expressed as follows,
		\begin{align}
		&h=y(\alpha_n^{\top} v_n+\zeta_n),\\
		&\mathrm{where} \hspace{4mm} v_{\iota+1}=y(\alpha_\iota^{\top} v_\iota+\zeta_\iota), \iota=1,2,...,n-1, \nonumber\\
		&\hspace{13mm} v_1=\phi(t),\nonumber
		\end{align}
		and $\alpha_\iota^{\top},\zeta_\iota$ are the weight and bias in the $\iota$th hidden layer, $y(\cdot)$ is the rectified linear unit activation function, and $\phi(t)$ is the state, which is the input of MLP. We build the actor process and critic process according to MLP.} 
	\subsection{Actor Process}
	We assume that the gradient policy $\pi_\theta(\phi(t),b_i(t))$ is differentiable in parameter $\theta$ and the update of $\theta$ is given as follows,
	\begin{align}
	\Delta \theta=\beta_a \nabla_\theta \mathcal{J}(\pi_\theta)=\beta_a \frac{\partial \mathcal{J}(\pi_\theta)}{\partial \pi_\theta} \frac{\partial \pi_\theta}{\partial \theta},
	\end{align}
	where $\beta_a$ is the learning rate for the actor and should be set small enough to avoid the oscillation of the policy \cite{mnih2016asynchronous}, because small updates of the value function will greatly influence the update of the policy. According to the maximum entropy principle \cite{maximumentropyprinciple}, we can utilize the Gaussian probability distribution \cite{sutton} to provide a parameterized policy to select continuous-charging action, which is represented as
	\begin{align}
	\pi_\theta(\phi(t),b_i(t))=\frac{1}{\sqrt{2\pi}\sigma}e^{-(b_i(t)-\varphi(\phi(t)))^2/2\sigma^2},
	\end{align}
	where $\varphi(\phi(t))$ is the average action value of this charging state, and $\sigma$ defines the standard deviation of all the possible charging actions. 
	Then $\pi_\theta(\phi(t),b_i(t))$ is the probability of choosing action $b_i(t)$ in state $\phi(t)$. According to $Q^{\pi_{\theta}}(\phi(t),b_i(t))$, we know the expected reward of the charging action $b_i(t)$ at state $\phi(t)$. Then we adjust the policy $\pi_{\theta}$ to make EV charging decisions.

	The objective of the actor-critic method is to find an optimal policy $\pi_{\theta}$ to maximize the following function,
	\begin{align}
	&\mathcal{J}(\pi_{\theta})\nonumber\\&=\mathbb{E}\{Q^{\pi_{\theta}}(\phi(t),b_i(t))\}\nonumber\\
	&=\int_{\Phi}\textcolor[rgb]{0,0,0}{D}^{\pi_{\theta}}(\phi(t))\int_{b_i(t)}\pi_{\theta}Q^{\pi_{\theta}}(\phi(t),b_i(t))\mathrm{d}b_i(t)\mathrm{d}\phi(t),
	\end{align}
	where $\textcolor[rgb]{0,0,0}{D}^{\pi_{\theta}}(\phi)$ is the state distribution function of policy $\pi_{\theta}$. We should optimize $\mathcal{J}(\pi_{\theta})$ by improving the parameters of policy $\pi_{\theta}$ iteratively. We utilize vector $\theta=(\theta_1,\theta_2,...,\theta_n)^{\top}$ to build the policy $\pi_\theta(\phi(t),b_i(t))=Pr(b_i(t)|(\phi(t),\theta))$. 
	\textcolor[rgb]{0,0,0}{In our implementation, the actor network has a fully-connected hidden layer with $200$ neurons, where state $\phi(t)$ is the input and parameter $\theta$ is the output. As mentioned above, $\pi_\theta(\phi(t),b_i(t))$ is the probability of choosing action $b_i(t)$ in state $\phi(t)$. The DNN can be trained to learn the best fitting parameter vector $\theta$ by iteratively minimizing the TD error.}
	\textcolor[rgb]{0,0,0}{\subsection{Critic Process}}
	The policy $\pi_{\theta}$ generates continuous actions from Gaussian distribution $\mathcal{N}(\mu_{\theta},\sigma_{\theta}^2)$. The expectation value $\mu_{\theta}$ is approximated by MLP.
	The temporal difference (TD) error is utilized to show the error between the approximation and the true value \cite{sutton}. TD error is defined as
	\begin{align}
	\label{deltat}
	\delta_t&= r(t+1)+\epsilon Q^{\pi_{\theta}}(\phi(t+1),b_i(t+1))-Q^{\pi_{\theta}}(\phi(t),b_i(t)),
	\end{align}
	where $r(t+1)$ is the reward in next time slot $t+1$, and $r(t+1)+\epsilon Q^{\pi_{\theta}}(\phi(t+1),b_i(t+1))$ is actual return following time $t$. 
	Similar to the parameters' update in the actor process, the parameters $\theta^v$ in critic process are updated as follows,
	\begin{align}
	\Delta \theta^v=\beta_c \delta_t \nabla_{\theta^v} Q^{\pi_{\theta^v}}(\phi(t),b_i(t)),
	\end{align}
	where $\beta_c$ is the learning rate for the critic and it should be chosen carefully because it will cause the oscillation if it is too large or it will take a long time to converge if $\beta_c$ is too small. \textcolor[rgb]{0,0,0}{In our implementation, the critic network has a fully-connected hidden layer with $100$ neurons, where state $\phi(t)$ is the input, and value function $Q^{\pi_{\theta^v}}(\phi(t),b_i(t))$ is the output.}

	\subsection{Actor-critic learning-based Smart Charging Algorithm}
\begin{figure} 
		\centering
		\includegraphics[width=0.48\textwidth]{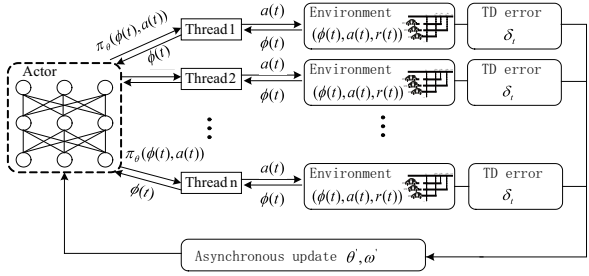}
		\caption{The architecture of SCA.}
		\label{A3Cnetwork}
	\end{figure}

	The complete description of actor-critic learning-based smart charging algorithm (SCA) is shown in Algorithm \ref{algorithm_EVCA3C}. The architecture of SCA is shown in Fig. \ref{A3Cnetwork}. \textcolor[rgb]{0,0,0}{First, we set the critic learning rate $\beta_c$, actor learning rate $\beta_a$, and the discount factor $\epsilon$. Then we have constructed the critic process and the actor process to develop SCA.}
	The critic process evaluates the policy from the state-action value function $Q^{\pi_{\theta^v}}(\phi(t),b_i(t))$ and the actor process has the following policy gradient, in which we use $\phi$ and $b$ as a logogram for state $\phi(t)$ and charging action $b_i(t)$,
	\begin{align}
	&\nabla_\theta \mathcal{J}(\pi_\theta)\nonumber\\&\approx \int_{\Phi}\textcolor[rgb]{0,0,0}{D}^{\pi_{\theta}}(\phi)\int_{b}Q^{\pi_\theta}(\phi,b)\nabla_\theta\pi_\theta(b|(\phi,\theta))\mathrm{d} b \mathrm{d}\phi
	\end{align}
	The actor parameter $\theta$ and critic parameter $\theta^v$ are updated simultaneously. To be more specific, the actor parameter $\theta$ is updated in the direction decided by the critic output. When actor-critic algorithm converges, the two sets of parameters are optimized.
	This algorithm has a policy $\pi(b_i(t)|(\phi(t),\theta))$ and an approximate value function of state $Q(\phi(t),\theta^v)$ and uses the multi-step returns \cite{sutton} to update the policy and the value function. The policy and the value function are updated after every $k_{\max}$ actions where the charging demands for EVs coming to the community are satisfied for an episode. According to Algorithm S3 in the reference \cite{mnih2016asynchronous}, the critic parameters are updated by follows,
	\begin{align}
	\label{omega1}
	d{\theta^v}'=d{\theta^v}'+\partial{\left(R(t)-Q(\phi(t);\theta^v)\right)^2}/{\partial \theta^v},
	\end{align}
	and the actor parameters are updated by follows,
	\begin{align}
	\label{theta1}
	d\theta'=d\theta'+\nabla_{\theta} \log\pi(b_i(t)|(\phi(t),\theta)) (R(t)-Q(\phi(t),\theta^v)),
	\end{align} 
	where the cumulative reward $R(t+1)=r(t)+\epsilon R(t)$ and the immediate reward $r(t)=-\sum_{i\in\mathcal{H}(t)}(k_0+2k_1b_i(t)+2k_1l_b(t))b_i(t)$. There is an agent in each thread, working in the copied environment. The gradient of one parameter is generated in each step. The gradients in many threads accumulate and parameters are shared and updated after the certain steps. \textcolor[rgb]{0,0,0}{After certain iterations, the reward will tend to converge, and the optimal charging solution will be achieved.}
	\allowdisplaybreaks[4]
	\begin{algorithm}
		\caption{Actor-critic Learning-based Smart Charging Algorithm (SCA)} 
		\label{algorithm_EVCA3C}
		\textbf{Input:} Critic learning rate $\beta_c$ and actor learning rate $\beta_a$, discount factor $\epsilon$, Gaussian policy $\pi_\theta(\phi(t),b_i(t))$, $b\sim N(\mu_{\theta},\sigma^2)$\\
		\textcolor[rgb]{0,0,0}{\textbf{Output:} Action $b_i(t)$}
		\begin{algorithmic}[1]
			\STATE \textbf{Initialization:} Thread step counter $t=1$, global shared counter $k=0$. Starting in state $\phi(0) \sim d^{\pi_\theta}(\phi(t))$, set parameter $\theta=\theta_0$ and $I=1$.
			\STATE \textbf{for} each thread \textbf{do}
			\STATE Reset gradients: $d\theta'=0$ and $d{\theta^v}'=0$.
			\STATE Synchronize thread parameters $\theta=\theta'$ and $\theta^v={\theta^v}'$
			\STATE Set $t_{start}=t$, get state $\phi(t)$
			\STATE \textbf{repeat}	
			\STATE \textbf{for} each step \textbf{do}
			\STATE {Select action $b_i(t+1)\sim \pi_\theta(\phi(t),b_i(t))$, move to next state $\phi(t+1)\sim P(\phi(t),b_i(t),\phi(t+1))$, then get immediate reward $r(t+1)$, update $k \leftarrow k+1$}
			\STATE \textbf{Critic:}	
			\STATE Update the basis function:\\ $\varPsi(\phi(t),b_i(t))=\nabla_\theta$ ln $\pi_\theta (\phi(t),b_i(t))$
			\STATE Update: $I=\epsilon I$\\
			$Q^{\pi_{\theta^v}}(\phi(t+1),b_i(t+1))={\theta^v}^{\top} \nabla_\theta$ ln $\pi_{\theta^v}(\phi(t+1),b_i(t+1))$ 	
			\STATE critic parameters: ${\theta^v}_{t+1}={\theta^v}_{t}+\beta_c\delta_t I$, where $\delta_t$ is updated by (\ref{deltat})
			\STATE \textbf{Actor:}
			\STATE Update the policy parameter: $\theta_{t+1}=\theta_t+\beta_a \delta_t\nabla_\theta J(\pi_\theta)$\\
			\STATE Update:	$\phi(t)\leftarrow \phi(t+1)$, $b_i(t)\leftarrow b_i(t+1)$, $z(t) \leftarrow z(t+1)$, $Q^{\pi_{\theta}}(\phi(t),b_i(t))\leftarrow Q^{\pi_{\theta}}(\phi(t+1),b_i(t+1))$
			\STATE \textbf{end \textcolor[rgb]{0,0,0}{for}}
			\STATE Perform asynchronous update of $\theta'$ using $d\theta'$ and of ${\theta^v}'$ using $d{\theta^v}'$ according to (\ref{omega1}), (\ref{theta1}).
			\STATE \textbf{until} $k>k_{\max}$\\
			\STATE \textbf{end \textcolor[rgb]{0,0,0}{for}}
		\end{algorithmic}
	\end{algorithm}
	
	\begin{figure} 
		\centering
		\includegraphics[width=0.48\textwidth]{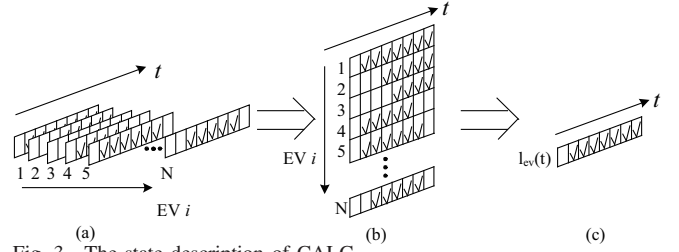}
		\vspace{-8mm}
		\caption{The state description of CALC.}
		\label{state_describe_improve}
	\end{figure}
	SCA takes the charging amount of each individual EV as the state and may suffer from high computational overhead. In the next section, we further develop a more computationally efficient customized algorithm by combining SCA with projection theorem, which takes the total charging amount as the state, and significantly reduce the state dimension.
	\subsection{Customized \textcolor[rgb]{0,0,0}{Actor-Critic Learning} Charging Algorithm}
	To be more computationally efficient, we further \textcolor[rgb]{0,0,0}{develop a customized \textcolor[rgb]{0,0,0}{actor-critic learning} charging algorithm (CALC) with two stages. In the first stage, the aggregate charging of EVs can be \textcolor[rgb]{0,0,0}{solved} by actor-critic learning. \textcolor[rgb]{0,0,0}{I}n the second stage, CALC finds a close-to-optimal charging schedule for each EV by \textcolor[rgb]{0,0,0}{the} projection theorem. Different from SCA that directly solves the charging amount for each EV, CALC takes the total charging amount of arriving EV fleet as the action and thus significantly reduces the dimension of the state space as shown in Fig. \ref{state_describe_improve}.} We see that the full state in Fig. \ref{state_describe_improve}(b) consists of charging actions for all $N$ EVs in Fig. \ref{state_describe_improve}(a), where the \textcolor[rgb]{0,0,0}{mark `$\surd$' indicates the available} time slots for charging. The reduced state, as shown in Fig. \ref{state_describe_improve}(c),  only considers the aggregate charging action instead of charging actions for each EV. 
	
	\textcolor[rgb]{0,0,0}{	
		In the first stage, we optimize the aggregate charging schedule $l_{ev}(t)=\sum\limits_{i\in\mathcal{H}(t)}b_{i}(t)$ by solving the following cost minimization problem:
		\begin{subequations} \label{cost_problem_projection} 
			\begin{align}
			&\min_{l_{ev}(t)}   \sum_{t\in\mathcal{T}}\Big(k_{0}l_{ev}(t)+2k_{1}l_{ev}(t)l_b(t)\Big)+k_{1}\Big(l_{ev}(t)\Big)^{2} \\
			&\rm{s.t.}\hspace{6mm} 0 \le l_{ev}(t) \le N_t b_{\max},
			\end{align}
		\end{subequations}
		where $N_t$ is the number of EVs in the charging station in time slot $t$, and $b_{\max}$ is the maximum \textcolor[rgb]{0,0,0}{charging amount in time slot $t$} of EV battery. Then we can get the optimal action $l^*_{ev}(t)$ by using the actor-critic learning. We have the state $\phi'(t)$ as follows,
		\begin{align}
		\phi'(t)=(SOC_{ev}(t),l_b(t)),
		\end{align}
		where $SOC_{ev}(t)=\sum\limits_{i\in\mathcal{H}(t)}SOC_i(t)$ is the total charging amount of EVs at the charging station in time slot $t$. We revise the reward function $r'(t)$ as follows.
		\begin{align}
		r'(t)=-(k_0+2k_1 l_{ev}(t)+2k_1l_b(t)) l_{ev}(t),
		\end{align}
		where the action $l_{ev}(t)=\sum\limits_{i\in\mathcal{H}(t)}b_i(t)$. We use actor-critic learning to get the \textcolor[rgb]{0,0,0}{optimal} charging solution $l_{ev}(t)$. Then we allocate the total charging amount $b_i(t)$ to each EV $i\in\mathcal{H}(t)$ by the CALC with Projection theorem according to the problem (\ref{cost_problem_projection}). 
		The charging station will adopt the close-to-optimal solution to Problem  (\ref{cost_problem_oa_discrete}) until a new EV comes in. When a new EV arrives, the profiles of $l_{ev}(t)$ will be updated.} 
	Note that the optimal aggregate charging amount $l^*_{ev}(t)$ derived in the first stage does not consider individual EV charging constraints in Problem (\ref{cost_problem}b) and thus may not be the optimal or even feasible for the offline problem in Problem (\ref{cost_problem}). Therefore, in the second stage, we aim to allocate a close-to-optimal charging schedule for each EV and make sure individual EV charging constraints are satisfied. Specifically, we solve the following projection problem,
	\begin{subequations} \label{problem_projection_b} 
		\begin{align}
		&\min_{b_{i}(t), b^*_i(t)} \sum_{t\in\mathcal{T}}\sum_{i\in\mathcal{H}(t)}||b_i(t)-b^*_i(t)||^2\\
		&\rm{s.t.} \it\sum_{t_{i}^{arr}}^{t_{i}^{dep}} b_{i}(t)=D_{i},i=1,2,...,N \\
		&\hspace{3mm} \it\sum_{i\in\mathcal{H}(t)} b^*_{i}(t)=l^*_{ev}(t)\\
		&\hspace{6mm} 0 \le b_{i}(t) \le b_{i,\max}\\
		&\hspace{6mm} 0 \le b^*_{i}(t) \le b_{i,\max}\\
		&\hspace{6mm} 0 \le L(t) \le L_{\max},
		\end{align}
	\end{subequations}
	where $b^*_i(t)$ is the temporary variable. We solve individual EV charging in the second stage, such that the charging allocation $b_i(t)$ is the closest to $b^*_i(t)$ of which the summation is the optimized aggregate charging schedule $l^*_{ev}(t)$ in the first stage. Therefore, we solve close-to-optimal solution for individual EV charging schedule that fulfills each EV's charging constraints.
	Using the reduced state, in stage-1, we optimize the total charging amount of all EVs $l^*_{ev}(t)$ without considering individual EV charging constraints. Note that the derived total charging amount may not be a feasible solution. Therefore, we construct the stage-2 problem, in which we reallocate the optimized total charging amount to each EV to fulfill each EV's charging constraints and thus obtain a close-to-optimal solution. \textcolor[rgb]{0,0,0}{Since CALC and SCA share the same core algorithm, i.e., the actor-critic learning, the convergence of CALC can be guaranteed when the learning rates $\beta_a$ and $\beta_c$ satisfy $\sum_{t=0}^{\infty}\beta_a=\infty$, $\sum_{t=0}^{\infty}\beta_c=\infty$ and $\sum_{t=0}^{\infty}\beta_a^2<\infty$, $\sum_{t=0}^{\infty}\beta_c^2<\infty$, according to \cite{Grondman2012}. Therefore, we need to set the learning rate properly for CALC to guarantee its convergence. For the computational complexity, please refer to \cite{Yan2019}. We also numerically show the convergence of our developed algorithms and their computational time in simulation results in Section \ref{sectionV}.}
	\section{Simulation} \label{sectionV}
	In this section, we evaluate the performance of three state-of-the-art algorithms, SCA and CALC by using practical load profiles.

	\subsection{Parameter-Settings for the Dynamic Simulation}
\begin{figure}[!ht]
		\centering
		\includegraphics[width=0.48\textwidth]{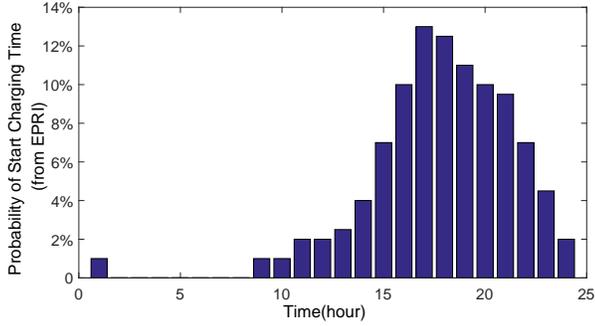}
		\caption{The probability of EVs' arrival over time \cite{Transportation2012}.}
		\label{result1}
	\end{figure}

\begin{figure}[!ht]
		\centering
		\includegraphics[width=0.48\textwidth]{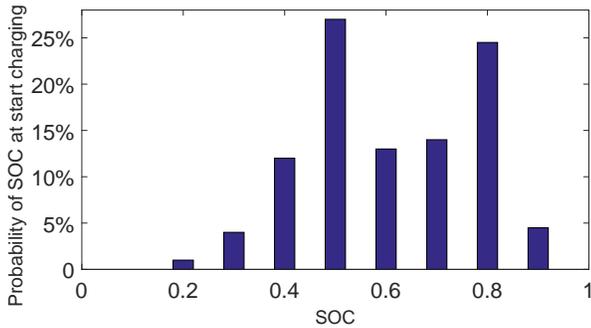}
		\caption{The probability of EVs' SOC at the arrival time \cite{Transportation2012}.}
		\label{result2}
	\end{figure}

	We adopt the base load profile in South California Edsion for two days from \cite{Gan2013}, that is, $T=48 h$. We set one time slot as $1 h$. \textcolor[rgb]{0,0,0}{The  arrival of EVs can be obtained from the statistical data in \cite{Transportation2012}. Fig. \ref{result1} shows the distribution of the EV arrival. The initial SOC of an  electric vehicle's battery affects the charging time and the load profile of the community. It is difficult to obtain the field measurement data and we can estimate the initial SOC value when EV arrives by a typical drive cycle, such as an urban dynamometer driving schedule \cite{UDDS}. Based on the measured statistics from \cite{ConsideringPowerSystemConstraintsandOperationCosts2016}, Fig. \ref{result2} depicts the probability density function of SOC of EVs' battery at their arrival.}
	This period is evenly divided into $48$ intervals.  We assume the same specifications for every EV in a scenario.
	We consider two types of EVs in two scenarios respectively \textcolor[rgb]{0,0,0}{based on two EV models \cite{iEV7L}: the first type has a maximal charging amount per time slot as $b_{\max}=3.2$ kW, and the battery capacity is $B_{\max}=36$ kWh. The other type has a maximal charging amount per time slot as $b_{\max}=1.4$ kW, and battery capacity is $B_{\max}=16$ kWh.}
	\textcolor[rgb]{0,0,0}{As discussed, our work models the distribution of the EV arrival, the probability density function about the SOC of EVs' battery at their arrival based on \cite{Transportation2012}. We generate the exogenous variables of EVs via Monte Carlo simulations using the distributions of arrival/departure patterns and charging demand. Note that if real-world measurements of EVs are available, they can be directly used in our algorithms as well.}
	We simulate these algorithms by Python on Win 10 x64. 
	
	\subsection{Performance Evaluation} 
	We evaluate the performance of SCA by using practical data and compare it with three benchmark algorithms as follows,
	
	1) Eagerly Charging Algorithm (EC)   \cite{OnlineCoordinatedChargingDecisionAlgorithmforElectricVehiclesWithoutFutureInformation}: EV $i$ draws the maximum amount of electricity \textcolor[rgb]{0,0,0}{from the charging station in each time slot. Thus, the charging amount in each time slot} is $b_{\max}$. We denote the cost by EC as $\varTheta_{EC}$. 
	
	2) Rolling online control algorithm (OA): EV $i$ draws the optimal power $b_{i}^*(t)$ from the charging station, which is the optimal value of Problem \ref{cost_problem_oa_discrete}. We denote the cost by OA as $\varTheta_{OA}$. 
	
	3) RL-based Adaptive Energy Management Algorithm (AEM)  \cite{ReinforcementLearningofAdaptiveEnergyManagement}: EV $i$ draws the power $b_{i}^*(t)$ from the charging station \textcolor[rgb]{0,0,0}{with the discrete-charging actions solved by the AEM algorithm, which is based on the Q-learning algorithm. AEM algorithm with discrete-charging actions is a good benchmark for our proposed algorithm with continuous-charging actions}. We denote the cost by AEM as $\varTheta_{RL}$. 
	
	\begin{figure}[!htbp] 
		\centering
		\includegraphics[width=0.48\textwidth]{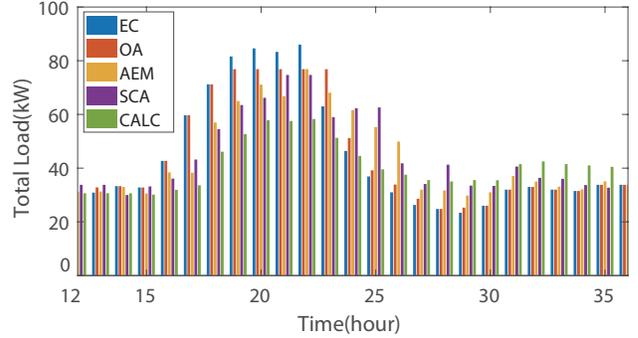}
		\vspace{-3mm}
		\caption{Comparison of total loads of $40$ Type-$1$ EVs.}
		\label{result_totalload5a_bar}
	\end{figure}%
	\begin{figure}[!htbp] 
		\centering
		\includegraphics[width=0.48\textwidth]{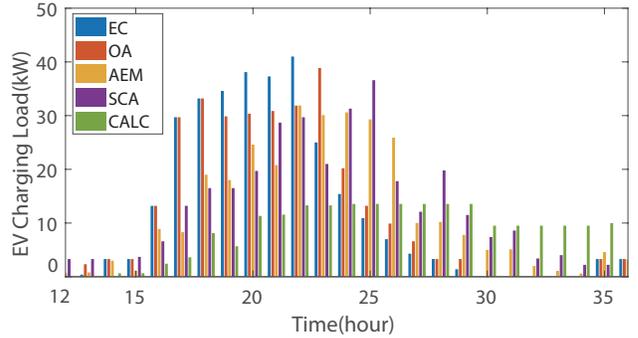}
		\vspace{-3mm}
		\caption{Comparison of EV charging loads of $40$ Type-$1$ EVs.}
		\label{result_EVload5a_bar}
	\end{figure}%
	
	
	Amongst all tested algorithms, SCA achieves the lowest total load peak. 
	We show the total loads about five algorithms of $40$ Type-$1$ EVs in Fig. \ref{result_totalload5a_bar} and EV charging loads about five algorithms of $40$ Type-$1$ EVs in Fig. \ref{result_EVload5a_bar}. \textcolor[rgb]{0,0,0}{From numerical simulations, we see that the peak loads are $86$ kW, $76.86$ kW, $76.9$ kW, $74.3$ kW, $58.6$ kW for of EC, OA, AEM, SCA, and CALC, respectively. CALC can reduce the peak load by $31.86\%, 23.76\%, 23.8\%, 21.13\%$, compared with EC, OA, AEM, and SCA.} 
	SCA and CALC have a lower load fluctuation than EC, OA and AEM algorithms and CALC has less load fluctuation than SCA.
	\begin{figure}[t]
		\centering
		\includegraphics[width=0.48\textwidth]{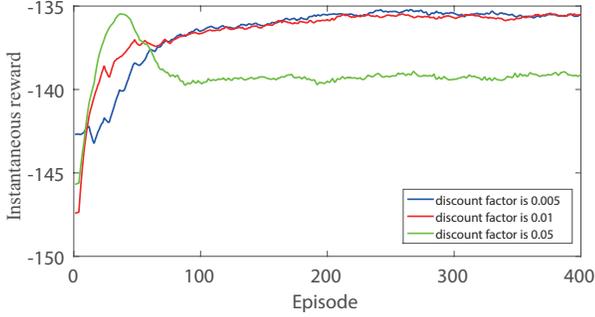}
		\caption{The total moving reward versus discount factor.}
		\label{rewardcostvsdiscountfactor}
	\end{figure}%
	\begin{figure}[!h]
		\centering
		\includegraphics[width=0.48\textwidth]{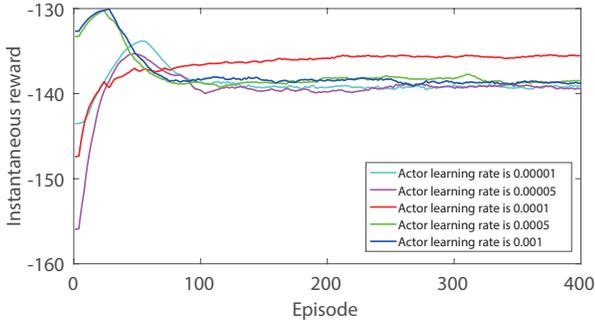}
		\caption{Total community load versus actor learning rate.}
		\label{rewardcostvslra}
	\end{figure}%
	\begin{figure}[!h]
		\centering
		\includegraphics[width=0.48\textwidth]{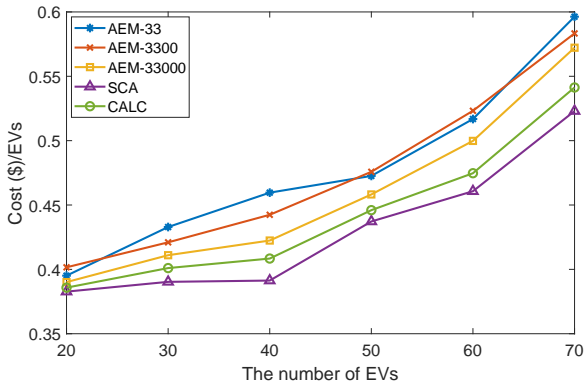}
		\caption{Comparison of average costs per EV of AEM, SCA, and CALC with different numbers of EVs.}
		\label{ratecostvsEV}
	\end{figure}%
	\begin{figure}[!h]
		\centering
		\includegraphics[width=0.48\textwidth]{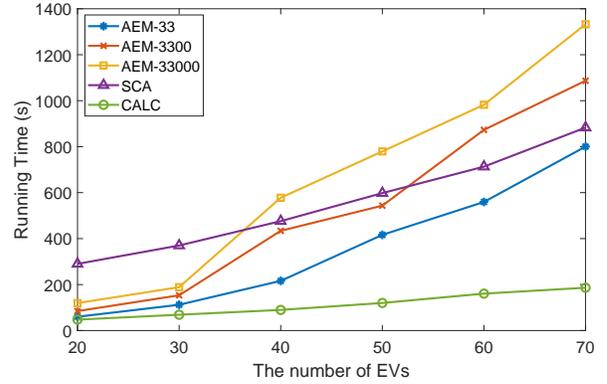}
		\caption{Comparison of the running time of AEM, SCA, and CALC during $48$ time slots with different numbers of EVs.}
		\label{runningtime_AEM_SCA}
	\end{figure}
	For Type-$2$ EV, the total EV charging costs of EC, OA, AEM algorithms and SCA are $\$15.14$, $\$13.47$, $\$12.67$, $\$11.51$ and the EV charging costs of EC, OA, AEM algorithms are $23.97\%$, $15.55\%$, $9.16\%$ higher than that of SCA, respectively. \textcolor[rgb]{0,0,0}{The EC algorithm is not an optimal algorithm where EVs are charged in a first-come-first-serve manner at the maximum rate. The OA algorithm solves the optimal EV charging over the rolling horizon but is not optimal in terms of long-term expected cost minimization. Q-learning method is used in the AEM algorithm with discrete-charging actions.} For Type-$1$ EV, the total EV charging costs of EC, OA, AEM algorithms, SCA and CALC are $\$28.67$, $\$19.45$, $\$18.22$, $\$16.01$, and $\$16.90$. The EV charging costs of EC, OA, and AEM algorithms are $24.03\%$, $21.49\%$, $13.80\%$ higher than that of SCA. \textcolor[rgb]{0,0,0}{We see that the performance of SCA is better than those of CALC and AEM, because CALC optimizes the aggregate charging schedule for the EV fleet in the first stage and finds a close-to-optimal charging schedule for each EV in the second stage.}

	From Fig. \ref{rewardcostvsdiscountfactor}, we can see that the discount factor can influence the convergence of SCA and CALC and it should be a very low value. The discount factor $\epsilon$ is an important parameter to reduce TD error in critic process. We simulate SCA under a set of discount factors $0.005,~0.01,~0.05$ and we can see that a low value can achieve a good performance. We can choose $\epsilon=0.01$ to balance the reward and convergence, achieving both reasonably fast convergence and high reward. Furthermore, the actor learning rate $\beta_a$ is also an important parameter to actor process. From Fig. \ref{rewardcostvslra}, we can see that the actor learning rate will influence the convergence of SCA and the update of policy. Large actor learning rates (e.g., $5 \times 10^{-4}$ and $10^{-3}$) lead to big overshoots of the rewards and the steady-state reward depends on the configuration of the actor learning rates as well. We see that $\beta_a = 10^{-4}$ achieves the best convergence performance and the highest reward among all the simulated rates. 
	
	\textcolor[rgb]{0,0,0}{Then we show the results of average costs per EV for all the tested algorithms in Fig. \ref{ratecostvsEV}. There are three levels of discrete-charging actions for AEM, where the charging action space is discretized into $33$, $3300$, and $33000$ even slices, respectively.} The average costs of all the tested algorithms steadily increase as the number of EVs increases. SCA achieves the lowest average cost among all simulated algorithms under all the scenarios of different numbers of EVs, validating the effectiveness of SCA.
	To show the time efficiency of SCA, we compare the performance of CALC with AEM algorithm and SCA and we can see that the total EV charging cost of CALC is $5.56\%$ higher than that of SCA \textcolor[rgb]{0,0,0}{and $7.24\%$ lower than that of AEM-33 algorithm, where the cost of CALC algorithm is $\$16.90$. When the charging action space is discretized into more slices, the charging cost will be lower.}
	The running time of SCA and CALC are $476$s and $90$s, of which scenario is $40$ EVs and $48$ time slots. The running time of SCA and CALC during $48$ time slots is shown in Fig. \ref{runningtime_AEM_SCA} and we can see that CALC has a significant higher time efficiency. \textcolor[rgb]{0,0,0}{Compared with SCA and CALC, AEM with more slices of discrete-charging actions has a larger computational complexity but a lower cost. When the charging action space is discretized into fewer slices, the computational complexity of AEM-$33$ is close to SCA. But if the charging action space of AEM is discretized into more slices, the computational complexity will increase rapidly as the number of EVs increases.}
	\section{CONCLUSION} \label{sectionVI}
	In this paper, we investigate an offline EV charging scheduling problem, which minimizes the charging cost of community EVs without future information. We reformulate an online optimization problem for EV charging and develop two actor-critic learning algorithms (namely SCA and CALC) supporting continuous-charging action. Based on our proposed SCA, we further develop a more computationally efficient CALC algorithm by reducing the state dimension and improving the computational efficiency. Simulation results show that SCA can outperform EC, OA, and AEM algorithms by $24.03\%$, $21.49\%$, $13.80\%$ in terms of energy cost, while achieving a good convergence. The total EV charging cost of CALC is $5.56\%$ higher than that of SCA but $7.24\%$ lower than that of AEM. CALC has a significantly higher computational efficiency and also achieves close-to-optimal performance compared with SCA. \textcolor[rgb]{0,0,0}{For our future work, we will consider reinforcement learning for the coordination of multiple charging stations. We will also consider vehicle-to-grid services as an emerging scenario for EV-grid interactions.}
	\bibliographystyle{IEEEtran}
	\bibliography{bibfile}
	
\end{document}